\documentclass[aps,preprint]{revtex4}%
\usepackage{amsmath}
\usepackage{graphicx}
\usepackage{subfigure}

\begin{document}
\preprint{ }
\title[ ]{Theoretical and experimental investigation of the light shift in Ramsey coherent population trapping}

\author{Yuichiro Yano}
\email{yano-yuichiro@ed.tmu.ac.jp}
\affiliation{Graduate School of Science and Engineering, Tokyo Metropolitan University, Minami-Osawa,
Hachioji, Tokyo 192-0397, Japan}

\author{Wujie Gao}
\affiliation{Graduate School of Science and Engineering, Tokyo Metropolitan University, Minami-Osawa,
Hachioji, Tokyo 192-0397, Japan}

\author{Shigeyoshi Goka}
\affiliation{Graduate School of Science and Engineering, Tokyo Metropolitan University, Minami-Osawa,
Hachioji, Tokyo 192-0397, Japan}

\author{Masatoshi Kajita }
\affiliation{National Institute of Information and Communications Technology, Koganei,
Tokyo 184-8795, Japan}


\begin{abstract}

The ac Stark shift (or light shift) of the 6$^{2}$S$_{1/2}$ (F $=$ 3 $\leftrightarrow $ 4) transition in $^{133}$Cs, as observed through coherent population trapping under  pulsed excitation, is measured using a $^{133}$Cs gas cell and the $D_1$-line vertical-cavity surface-emitting laser.
This light shift can be calculated using density-matrix analysis.
We derive an expression for this shift as a function of light intensity, showing that it varies linearly with respect to light intensity only with intensities higher than 1.0~mW/cm$^2$.
For pulsed excitation of high laser intensity, the variation in light shift is 20 times lower than that when using a continuous wave.
The differences between the results of theory and experiment are discussed, taking into account the difference in conditions assumed; the results from theoretical analysis, taking the attenuation of the first-order sideband into account, approximately agree with the experimental results.
The light shift is reduced by shortening the observation times.
\end{abstract}

\startpage{1}
\endpage{2}
\maketitle

\section{Introduction }
Compact atomic clocks are required in many fields of applications such as telecommunications, navigation systems, and synchronization of networks \cite{ct1}.
As compact frequency references, atomic clocks based on coherent population trapping (CPT) have attracted much attention \cite{ct2}.
In general, atomic clocks are wanted for their long-term frequency stability.
Instabilities arise owing to changes in the measurement conditions.

The buffer gas and light shifts in CPT atomic clocks determine the limiting factors in long-term stability \cite{ct3}.
The buffer gas is often contained in the gas cell to suppress the relaxation of atoms owing to collisions between them and the vapor cell wall, which is important for observing the narrow resonance linewidth. 
However, the frequency shift is caused by and depends on the temperature of the gas cell \cite{ct4}.
Fortunately, combining two buffer gases having opposing frequency shifts with temperature changes does reduce the temperature dependence \cite{ct5,ct6} and enables a small temperature coefficient of frequency (TCF) to be obtained.
Using a Ne buffer gas, a zero TCF has recently been reported in the 70 $^\circ$C to 80 $^\circ$C range \cite{ct7}.
The light shift is the Stark frequency shift induced by laser light.
The light shift under a continuous wave (cw) increases linearly with increasing light intensity.
Because the sensitivity of the frequency to the light intensity is greater than that from temperature, the light shift is a dominant factor limiting the long-term stability of CPT atomic clocks \cite{ct8}. 

Several methods have been proposed to suppress the light shift \cite{ct9,ct10,ct11}. One particular method previously reported is the Raman-Ramsey scheme \cite{ct12,ct13,ct14}.
A high-frequency stability of 3$\times$ 10$^{-14}$ at 200 s can be obtained using a Cs vapor cell \cite{ct15}.
However, the relationship between the light shift and the pulse parameters is still unclear.
The light shift dependencies on pulse parameters, such as the light intensity, free evolution time, and observation times, need to be studied to enhance frequency stability for longer durations.

In this study we investigate both theoretically and experimentally the light shift of the 6$^{2}$S$_{1/2}$ (F $=$ 3 $\leftrightarrow $ 4) transition frequency in $^{133}$Cs under pulse excitation.
We calculate the light shift under pulse excitation based on density-matrix analysis and measure the light shift using a $^{133}$Cs gas cell and the $D_1$-line vertical-cavity surface-emitting laser (VCSEL).
First, we derive the calculated and measured light shifts as a function of light intensity.
The results show that this shift is significantly less than that under a cw; it also varies nonlinearly with respect to light intensity.
Second, we derive the light shift as a linear function of the observation time.
Finally, we discuss the differences in conditions assumed in the theory and from experiment.
The quantitative discrepancy between calculated and experimental results stems from the difference in these conditions.

\section{Theory}
Figure 1(a) shows the excitation scheme using a left circularly polarized $\sigma ^{+}$  light field for the $^{133}$Cs-$D_1$ line.
In CPT phenomenon, two ground states of 6$^{2}$S$_{1/2}$ are coupled simultaneously to a common excited state of 6$^{2}$P$_{1/2}$. In this system, the dynamical behavior of the density matrix $\rho $ is governed by the quantum Liouville equation,
\begin{equation}
\label{eq1}
\frac{\partial}{\partial t}\rho (t)=\frac{i}{\hbar }[\rho ,H]+R\rho,
\end{equation}
where $H$ is the Hamiltonian matrix for this three-level system and $R$ stands for the relaxation terms. Using the rotating-wave approximation with the simplified $\Lambda $-type model depicted in Fig. 1(b), Eq.(\ref{eq1}) can then be rewritten as 

\begin{eqnarray}
\begin{split}
\dot{\rho}_{11}&=i\frac{\Omega_p}{2}(-\rho_{13}+\rho_{31})+\Gamma_{31} \rho_{33}+\gamma_s(\rho_{22}-\rho_{11}), \\
\dot{\rho}_{22}&=i\frac{\Omega_c}{2}(-\rho_{23}+\rho_{32})+\Gamma_{32} \rho_{33}-\gamma_s(\rho_{22}-\rho_{11}), \\ 
\dot{\rho}_{33}&=i\frac{\Omega_p}{2}(\rho_{13}-\rho_{31})+i\frac{\Omega_c}{2}(\rho_{23}-\rho_{32})-\Gamma_3 \rho_{33}, \\
\dot{\rho}_{12}&=i \rho_{12}(\delta_p-\delta_c)-i\frac{\Omega_c}{2}\rho_{13}+i\frac{\Omega_p}{2}\rho_{32}-\gamma_s \rho_{12}, \\
\dot{\rho}_{13}&=i\rho_{13}\delta_p-i\frac{\Omega_c}{2}\rho_{12}+i\frac{\Omega_p}{2}(\rho_{33}-\rho_{11})-\gamma_f \rho_{13}, \\
\dot{\rho}_{23}&=i\rho_{23}\delta_c-i\frac{\Omega_p}{2}\rho_{21}+i\frac{\Omega_c}{2}(\rho_{33}-\rho_{22})-\gamma_f \rho_{23}, 
\end{split}
\label{eq2}
\end{eqnarray}
\noindent where $\Omega_p$ and $\Omega_c$ are Rabi frequencies and $\delta_p$ and $\delta_c$ are frequency detunings, with the trace of the density matrix satisfying the closed-system condition
\begin{equation}
\label{eq3}
\rm{Tr}(\rho )=\rho_{11} +\rho_{22} +\rho_{33} =1.
\end{equation}

\noindent Here $|1\rangle$ and $|2\rangle$ correspond to the two ground states $|F = 3, m_{F} =0\rangle$ and $| F = 4, m_{F} = 0 \rangle$ in the 6$^{2}$S$_{1/2}$ state and $|3\rangle$ corresponds to a state in 6$^{2}$P$_{1/2}$.

The total emission rate is $\Gamma_3=\Gamma_{31}+\Gamma_{32}$, $\Gamma_{31}=\Gamma_{32}$, $\gamma_f=\Gamma_3 /2$, and the ground-state relaxation rate $\gamma_s$ is a minuscule quantity, $\gamma_s \ll \gamma_{f}$.
In this calculation, the total emission rate was set at 370 MHz, which was obtained experimentally from absorption lines of Cs cell with Ne buffer gas.
The obtained emission rate of 370 MHz is consistent with the value estimated by Ref. \cite{ct23}.
The ground-state relaxation rate $\gamma_s$ was set to zero because $\gamma_s$, being smaller than 300 Hz, does not alter the calculation result.
For bichromatic light, we have the relation $\delta_p=-\delta_c=\Delta_0/2$.

Figure 2 shows the Ramsey interrogation sequence \cite{ct17}.
The first laser pulse of duration $\tau$ irradiates the atoms into the dark state. After a free evolution time $T$, a second pulse irradiates the atoms again.
Immediately after the pulse rise, the transmitted intensity of the second pulse is measured after time $\tau _{o}$ and the atoms are prepared for the next measurement.
This pulse cycle is repeatedly executed until the atomic state reaches equilibrium. 
The Rabi frequency for this scheme is written

\begin{eqnarray}
\label{eq4}
\begin{split}
 &\Omega_{p} ,\Omega_{c} >0&(&0<t<\tau ), \\ 
 &\Omega_{p} =\Omega_{c} =0&(&\tau <t<\tau +T). \\ 
 \end{split}
\end{eqnarray}

As the Liouville equation does not contain the frequency shift term with regard to the light intensity, and the light field will shift the energy levels of the two ground states via the light shift, we include the light shift terms in Eq. (\ref{eq2}):

\begin{eqnarray}
\begin{split}
\label{eq5}
 \delta_{p} &=&\frac{\Delta_0}{2}- {\mathcal S}_1(&\quad\,\frac{\Delta_0}{2},\Omega_{p}), \\ 
 \delta_{c} &=&-\frac{\Delta_0}{2}- {\mathcal S}_2(&-\frac{\Delta_{0}}{2},\Omega_{c}), 
\end{split}
\end{eqnarray}
where $\mathcal{S}_1$ and $\mathcal{S}_2$ are the light shifts  under a cw. These light shift terms  includes the light shift arising from the higher-order sidebands (see the Appendix).

The density matrix element $\rho_{33}$, describing the population of the excited state, is proportional to the fluorescence intensity. When the atoms fall into the dark state, fluorescence is at a minimum. We therefore investigate $\rho_{33}$ as a means to calculate the light shift. Figure 3 shows $\rho_{33}$ as a function of the Raman detuning $\Delta_{0}$ using Eq. (\ref{eq2}) with Eq. (\ref{eq5}). We can see that the line shape of the Ramsey CPT fringe can be calculated from Fig. 3(a). Figure 3(b) shows the center part fringe of Ramsey CPT. The solid line is calculated using Eq. (\ref{eq5}), whereas the dotted line is calculated without using Eq. (\ref{eq5}). The Ramsey fringe exhibits a slight asymmetric resonance with respect to Raman detuning. Clearly, the minimum value of $\rho_{33 }$ calculated using Eq. (\ref{eq5}) is shifted.

\section{Experimental setup}
The experimental setup is shown in Fig. 4. The measurement system is based on a previous Ramsey-CPT observation system \cite{ct13}.

A single-mode VCSEL (Ricoh Company Ltd., Japan) was used to simultaneously excite two ground states to the common excited state.
The wavelength of the VCSEL used to excite $^{133}$Cs at the $D_1$-line was 895 nm.
The VCSEL was driven by a dc injection current using a bias $T$ and was modulated at 4.6 GHz using an analog signal generator to generate the first-order sidebands around the laser carrier. 

For pulse excitations, an acousto-optical modulator (AOM) was used to modulate the light intensity.
The AOM has a nominal rise and fall time of 65 ns. The total light intensity incident on the gas cell was adjusted using the control voltage of the AOM and was calibrated by the optical power meter.
A Pyrex gas cell was used that contained a mixture of $^{133}$Cs atoms and Ne buffer gas at a pressure of 4.0 kPa.
This cell was cylindrical with a diameter of 20 mm and optical length of 22.5 mm.
Its temperature was maintained at 42.0 $^\circ$C. The gas cell and Helmholtz coil were covered with a magnetic shield to prevent external magnetic fields from affecting the magnetic field inside the cell.
The Helmholtz coil produced the internal magnetic field in the gas cell.
The axis of a 10 $\mu $T magnetic field was set parallel to the direction of the laser light ($C$-axis direction).

\section{Results}
\subsection{Light shift as a function of the light intensity}
\label{sec:section}
Figure 5 shows the light shift under pulse excitation as a function of intensity.
The dashed line is the calculated light shift under pulse excitation.
The values of the sideband distribution are used in Fig. 7 (shown in the Appendix).
These parameters are obtained by measuring the absorption spectrum of the Cs gas cell.
The calculated variation of the light shift under a cw is 30.9 Hz/(mW/cm$^2)$ and the light shift under a cw is proportional to the light intensity over all ranges of light intensity. 
The light shift under pulse excitation increases with increasing light intensity and is significantly less than that under a cw.
Note that under pulse excitation the light shift depends nonlinearly on light intensity and, stemming from its linear behavior, there is a threshold intensity associated with the variation in shift.

For small light intensities ($<$ 1.0 mW/cm$^2$), the light shift varies nonlinearly with light intensity, and the variation in light shift decreases with increasing light intensity.
The maximum differential light shift value is estimated at 14.0 Hz/(mW/cm$^{2})$ with the absence of light intensity.
Because the maximum differential value under pulse excitation is less than that under a cw, pulse excitation enables a lower sensitivity of frequency to light intensity compared with that under a cw.
For high light intensities ($\geq$ 1.0 mW/cm$^2$), the light shift linearly increases with increasing light intensity.
The straight line is the fitting curve of the light shift in the intensity range from 1.0 to 2.5 mW/cm$^{2}$.
Under pulse excitation, the variation in light shift is 0.594 Hz/(mW/cm$^{2})$, which is 50 times smaller than that under a cw.
Because this variation is significantly smaller than that for low light intensity, to reduce the light intensity sensitivity the atoms need to be excited using high light intensities.

The same trend is also demonstrated from the experimental results plotted by the dots.
The resonance frequency at which light intensity is zero is 4.592,325 GHz shifted from the unperturbed hyperfine transition frequency by 18 kHz.
The frequency shift is caused by the buffer gas shift. 
The measured variation in light shift under a cw is 28.4 Hz/(mW/cm$^{2})$, which is almost consistent with the calculated value.
However, the measured light shift under pulse excitation is three times larger than that from the calculation with an intensity higher than 1.0 W/cm$^2$.

The most likely cause of the difference between calculation and experiment is the attenuation of the first-order sidebands in the gas cell.
The first-order sidebands incident on the gas cell are absorbed by interacting Cs atoms through the gas cell and the absorbed energy is converted to luminescence or kinetic energy of the Cs atoms.
Therefore, for the gas cell, the light intensities of the first-order sidebands decrease exponentially. 

Since the sideband distribution varies depending on the position in the gas cell,
the light shifts under a cw $\mathcal{S}_1$, $\mathcal{S}_2$ and the Rabi frequencies $\Omega_p$ and $\Omega_p$ also depend on the position in the gas cell.
In this calculation, for the sake of simplicity, the intensity ratio between the actual light intensity and the calculated light intensity is defined as $r$ ($=I_{p'}/I_p=I_{c'}/I_c$).
The $I_{p'}$ and $I_{c'}$ are actual light intensities contributing to excite atoms. 
The light shift taking into account the absorption effect for the first-order sidebands is plotted by the solid line in the Fig. 5.
The intensity ratio $r$ is set equal to 0.5 because its actual mean intensity is measured at 50\%.
The measured value is experimentally obtained from the transmitted light intensity through the gas cell.
It is clearly shown that the calculated curve with $r=0.5$ is close to that of the experimental data.

\subsection{Light shift as a function of observation time}
Figure 6 shows the light shift as a function of the observation times $\tau_{o}$ under different light intensities. 
The dashed line is the calculated light shift under pulse excitation without taking into account the absorption effect for the first-order sideband, and the solid line is the calculated light shift taking into account the absorption effect.
The calculation is performed with the same conditions as stated in Sec. \ref{sec:section}.
For short observation times ($<$ 200 $\mu $s), the light shift linearly increases with increasing observation times $\tau_{o}$, with
the variation in light shift increasing as well. Hence, a smaller light shift can be obtained when setting a shorter $\tau_{o}$.
For long observation times ($\geq$ 200 $\mu $s), the light shifts are saturated and the obtained values are constant regardless of the observation time $\tau_{o}$.
The saturation time becomes shorter with higher light intensity.
Because the frequency shift at the saturation time is almost equal to the light shift under a cw, the atoms fall into a dark state under a cw and fall more quickly into the dark state for higher light intensities because the rate of pumping to steady CPT increases with increasing light intensity.

The dots refer to the measured light shift.
The experimental results show the same trend as the calculation results.
However, the difference between the experimental and calculation results ($r= 1.0$) occurs in the light shifts and the saturation time $\tau_{s}$.
The measured saturation time is about two times longer than the calculated time in the case of $r=1.0$.

Taking into account the absorption effect, the calculated curves plotted by the solid line are closest to that of the experimental data.
Because of the reduced pumping rate, the saturation time is longer than that without considering the first-order sidebands attenuation.
The relative error between calculation and experiment is less than 10 \%, which corresponds to the relative frequency difference under a cw between calculation and experiment.
These results infer that quantitatively the attenuation from the first-order sidebands is required in the calculation of the light shift under pulse excitation.

The calculation curve is close to the experimental one; however, the residual frequency between calculation and experiment remains.
Since the residual frequency increases with increasing light intensity,
one reason for the residual is that the actual light intensity rate $r$ is unsuitable for the approximation of first-order sideband attenuation under high light intensity because the actual light intensity is absolutely away from the incident light under high light intensity.
Therefore, this implies that the calculation with spatial decomposition will be needed for further quantitative calculation.

\section{Conclusion}
We investigated both theoretically and experimentally the light frequency shift of the 6$^{2}$S$_{1/2}$ (F $=$ 3 $\leftrightarrow $ 4) transition in $^{133}$Cs under pulse excitation.
We calculated the light shift in the Ramsey CPT based on density-matrix analysis and measured the light shift using a $^{133}$Cs gas cell and the $D_1$-line VCSEL.
We found that the variation in light shift under pulse excitation is 20 times lower than that under a cw, noting that the light shift is nonlinear in intensity.
It was found necessary to excite the atoms using a high light intensity to reduce the light intensity sensitivity.
We discussed the difference between conditions assumed in the theoretical and experimental evaluations.
There is good agreement between the calculation and experiment when taking the attenuation of the first-order sidebands into account.

We also derived an expression for the light shift as a function of the observation time.
The light shift under pulse excitation is dependent on the observation times and increases with increasing observation time.
The results indicated a smaller light shift when setting a shorter $\tau_o$.

\section{Appendix}
In this appendix, we derive the light shift for two ground states, $\mathcal{S}_1$ and $\mathcal{S}_2$, under a cw. The light shift contains the frequency shift taking into account the higher-order sidebands emitted by the VCSEL.

Because the Rabi frequency is smaller than the natural linewidth, the light shift can be calculated ignoring the Autler-Townes effect\cite{ct22}.
Therefore, the light shift between the ground and excited states can be simply calculated as \cite{ct19}
\begin{equation}
\label{eq6}
\mathcal{S}=\frac{1}{4} \frac{\Omega^2 \Delta }{\Delta^2+ \Gamma^2/4},
\end{equation}

\noindent where $\Omega$ is the Rabi frequency, $\Delta $ is the detuning from the absorption line, and $\Gamma$ is the total emission rate.
The VCSEL is generally used as the light source for the CPT atomic clocks and the injection current of the VCSEL is modulated by a rf generator to generate the two coherent laser beams around the laser carrier.
However, the modulation generates higher-order sidebands together with the two first-order sidebands.
As the higher-order sidebands are located near the absorption line, the light shift is induced by the higher-order sidebands as well as the two first-order sidebands.

In the frequency domain, the $n$th-order sideband intensity $I(n)$ is defined as the intensity within $\omega_{0} + n\omega_{m}$.
Figure \ref{fig:sideband} shows the normalized sideband distribution irradiated from VCSEL in the $n$th-order range from $-$3 to 3.
The sideband distribution is obtained from absorption lines of the Cs cell without the buffer gas and the measurement method is based on Ref. \cite{ct21}.
Higher-order sidebands are also generated together with the two first-order sidebands and the sideband intensity decreases with increasing order $n$.
In addition, as the higher-order sideband's detuning from the absorption line increases with increasing $n$, the influence on the light shift is small as $n$ increases.
Therefore, this implies that the terms in the light shift using higher-order sidebands over 4 ($|n|>3$) can be ignored.
The electric field of the $n$th-order sideband is calculated to be
\begin{equation}
\label{eq10}
E(n)=\sqrt{\frac{2I(n)}{\varepsilon_0 c}}.
\end{equation}

In the $^{133}$Cs $D_1$ line, the Rabi frequency for each sideband is
\begin{eqnarray}
\label{eq11}
\Omega_{ij}(n)&=&\frac{\langle i|\hat{\mu} E(n)|j\rangle}{\hbar}\\ \nonumber
&=&\frac{E(n)}{\hbar}\langle i|e_q \hat{r}|j\rangle =\sqrt{\frac{2I(n)}{\varepsilon_0 c \hbar^2}}\langle i|e_q \hat{r}|j\rangle.
\end{eqnarray}

\noindent Here $| i \rangle$ is the ground state 6$^{2}$S$_{1/2}$ and $| j \rangle$ is the excited state of 6$^{2}$P$_{1/2}$. In particular, since $n = 1$ and $-1$ sidebands are components exciting two ground states, $\Omega_{p}$ and $\Omega_{c}$ correspond to $\Omega_{34'}$ and $\Omega_{44'}$.
The detunings $\Delta_{ij}(n)$ of the $n$th-order sidebands are
\begin{eqnarray}
\Delta_{ij}(n)= \left\{ 
\begin{split}
&\Delta_{44'}(n)=\delta_c +\frac{f_S}{2}(n+1),\\
&\Delta_{43'}(n)=\Delta_{44'}(n) +f_P,\\
&\Delta_{34'}(n)=\delta_p +\frac{f_S}{2}(n-1),\\
&\Delta_{33'}(n)=\Delta_{34'}(n)+f_P,\\
\end{split}
\right.
\label{eq12}
\end{eqnarray}
where $f_S$ is the frequency difference between the two ground states of 6$^2$S$_{1/2}$ (F $=$ 3 $\leftrightarrow $ 4) and $f_P$ that between the two excited states of 6$^2P_{1/2}$(F' $=$ 3 $\leftrightarrow $ 4).

From Eqs. (\ref{eq11}) and (\ref{eq12}), the light shifts $\mathcal{S}_1$ and $\mathcal{S}_2$ of the two ground states can be calculated using Eq. (\ref{eq6})
\begin{eqnarray}
\begin{split}
\mathcal{S}_1=\sum^{3}_{n=-3}\frac{1}{4}\frac{\Omega_{34'}^2(n)\cdot \Delta_{34'}(n)}{\Delta_{34'}^2(n)+\Gamma_{4'}^2/4}+\frac{1}{4}\frac{\Omega_{33'}^2(n)\cdot \Delta_{33'}(n)}{\Delta_{33'}^2(n)+\Gamma_{3'}^2/4} ,\\
\mathcal{S}_2=\sum^{3}_{n=-3}\frac{1}{4}\frac{\Omega_{44'}^2(n)\cdot \Delta_{44'}(n)}{\Delta_{44'}^2(n)+\Gamma_{4'}^2/4}+\frac{1}{4}\frac{\Omega_{43'}^2(n)\cdot \Delta_{43'}(n)}{\Delta_{43'}^2(n)+\Gamma_{3'}^2/4}. 
\end{split}
\label{eq13}
\end{eqnarray}

We then obtain the light shift under a cw using the modulated VCSEL:
\begin{equation}
\mathcal{S}_{12}=\mathcal{S}_2-\mathcal{S}_1
\label{eq14}
\end{equation}

Figure \ref{fig:light shift_CW} shows the light shift as a function of the light intensity. 
The constant data for Cs are used in Ref. \cite{ct18}.
The light shift is proportional to the light intensity. 

\section*{Acknowledgments}
The authors are grateful to Ricoh Company Ltd. for providing us with the Cs-D$_1$ VCSEL.

\section{Figure captions}
\begin{figure}[h]
  \centering
    \includegraphics[width=4in]{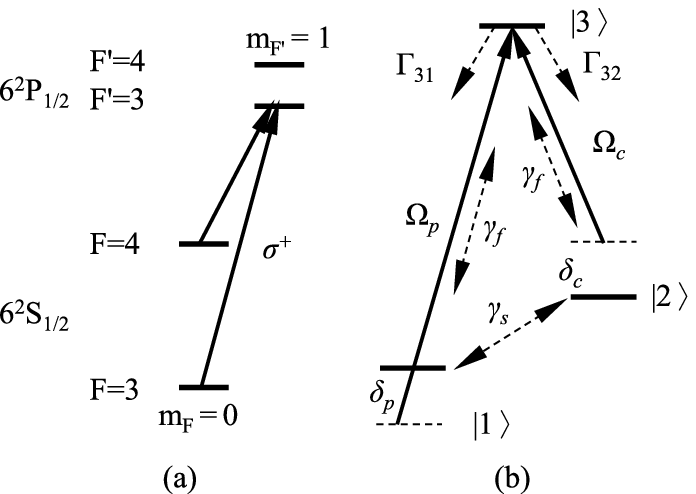}
  \caption{(a) Excitation scheme with a left circularly polarized $\sigma^+$ light field on the $D_1$ line of Cs. (b) Closed $\Lambda $-type three-level model used to calculate CPT phenomenon: $\delta_p$ and $\delta _c$ are detunings of the probe laser and coupling laser, respectively; $\Omega_p$ and $\Omega_{c}$ are Rabi frequencies; $\Gamma_{31}$ and $\Gamma _{32 }$ are the relaxation terms between an excited state and the two ground states, respectively; $\gamma_{f}$ is the decoherence rate between the excited state and ground states; and $\gamma_{s}$ is the decoherence rate between the two ground states}
\end{figure}

\begin{figure}[h]
  \centering
    \includegraphics[width=4in]{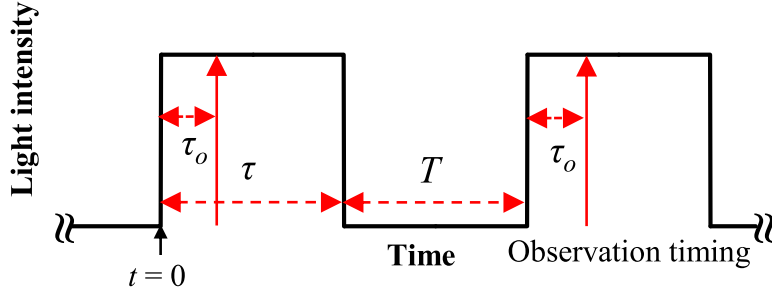}
  \caption{Ramsey pulse sequence: $\tau $ is the excited time, $T$ the free evolution time, and $\tau_{o}$ the observation time of the resonance signal.
}
\end{figure}

\begin{figure}[h]
  \centering
  \subfigure[]{
    \includegraphics[width=4in]{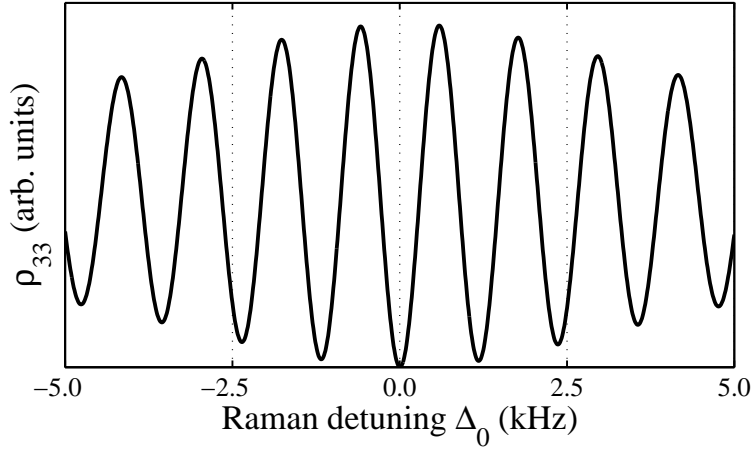}
    }
    \\
  \subfigure[]{
    \includegraphics[width=4in]{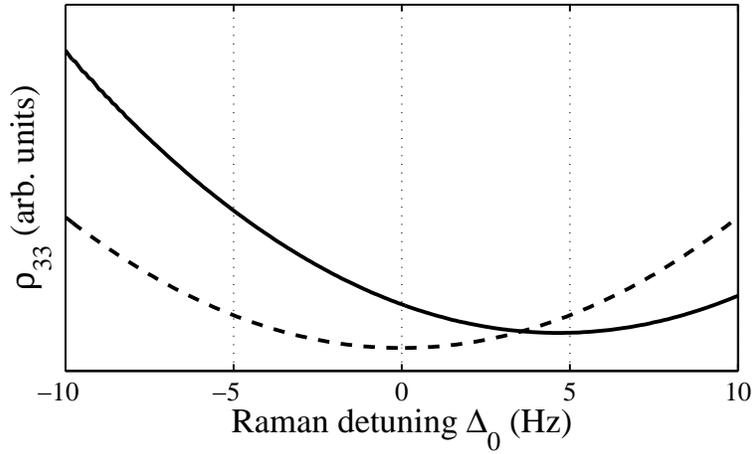}
    }
  \caption{(a) Line shape of $\rho_{33}$ in the detuning range from $-$5.0 to 5.0 kHz. (b) Line shape of $\rho_{33}$ in detuning range from $-$10 to 10 Hz. The solid (dotted) line is calculated with (without) Eq. (\ref{eq5}). The excitation duration time $\tau $ is 800 $\mu $s and the free evolution time $T$ is 800 $\mu$s. The observation time $\tau_{o}$ is 10 $\mu $s. The emission rate is 370 MHz . The total light intensity is 2.5 mW/cm$^2$. The sideband distribution is used in Fig. 7 and shown in the Appendix}
\end{figure}

\begin{figure}[h]
  \centering
    \includegraphics[width=4in]{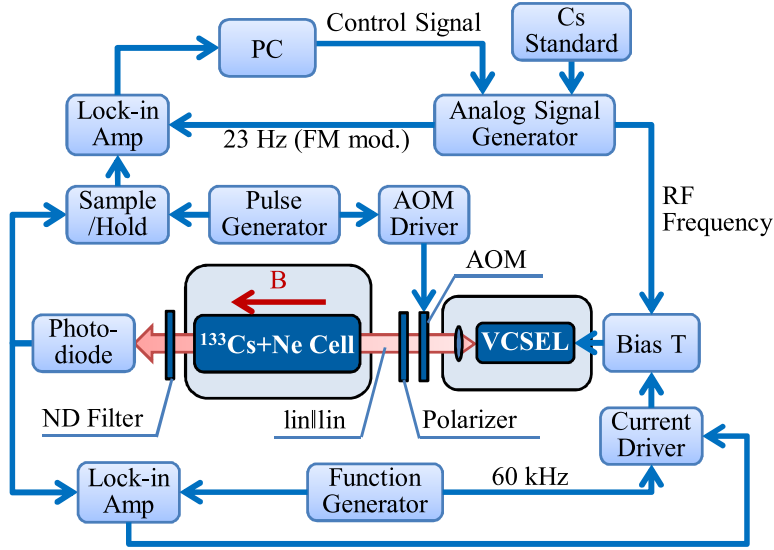}
  \caption{Experimental setup. PC: Personal computer, ND Filter: Neutral density filter.}
\end{figure}

\begin{figure}[h]
  \centering
    \includegraphics[width=4in]{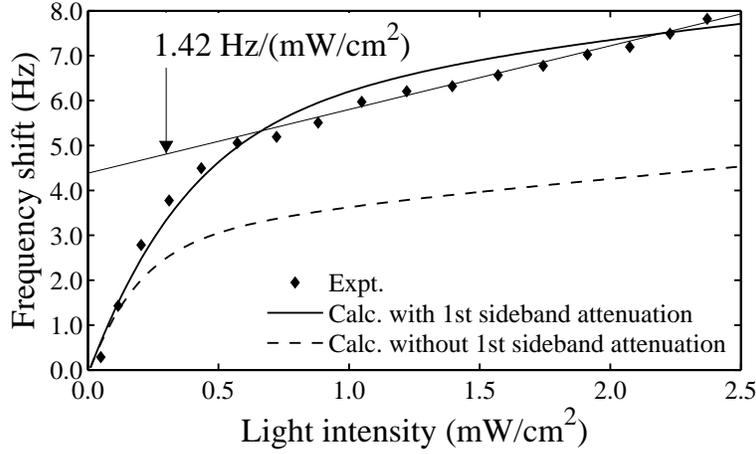}
  \caption{Light shift as a function of light intensity under pulse excitation taking into account the absorption effect for the first-order sidebands, assuming that their actual mean intensities are 50 \% of the incident light. The dots are experimental data. The solid (dashed) line is the light shift under pulse excitation with (without) the attenuation of the first-order sidebands. The excitation duration time $\tau$ is 800 $\mu$s and the free evolution time $T$ is 800 $\mu$s. The observation time $\tau_{o}$ is 10 $\mu$s.}
\end{figure}

\begin{figure}[p]
    \includegraphics[width=4in]{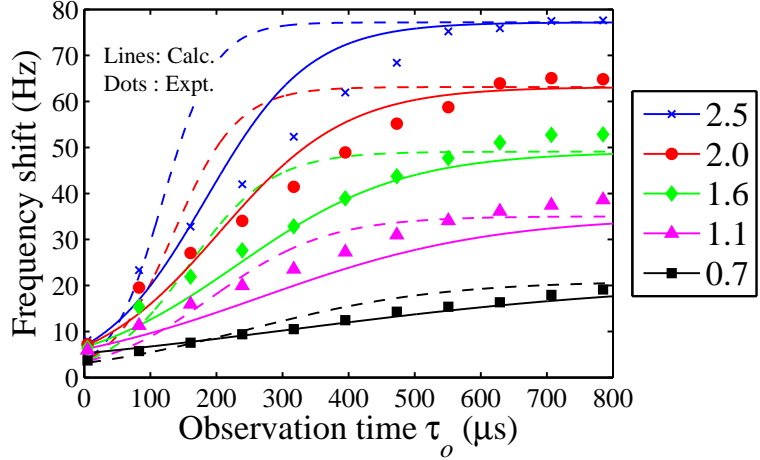}
  \caption{Light shift as a function of observation time $\tau_o$ under different light intensities (mW/cm$^2$) taking into account the absorption effect for the first-order sidebands, assuming that their actual mean intensities are 50 \% of the incident light. The dots are experimental data. The solid (dashed) line is the light shift under pulse excitation with (without) the attenuation of the first-order sidebands. The excitation duration time $\tau$ is 800 $\mu$s and the free evolution time $T$ is 800 $\mu$s.}
\end{figure}

\begin{figure}[p]
  \centering
    \includegraphics[width=4in]{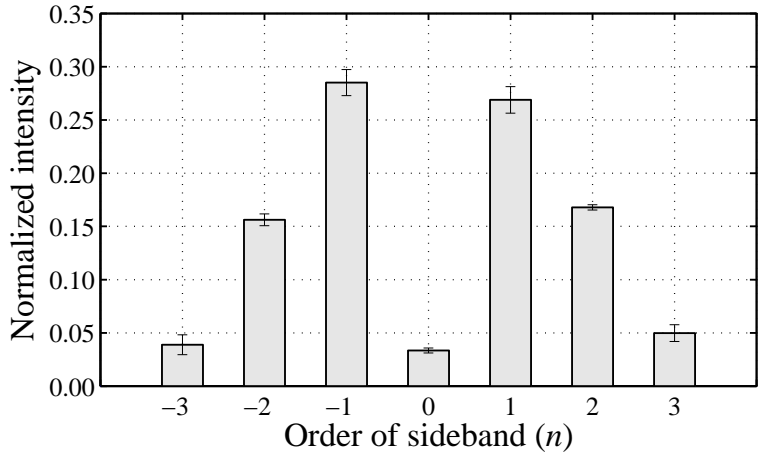}
  \caption{Normalized sideband intensities estimated based on Ref. \cite{ct21}. Error bars represent standard deviations.}
    \label{fig:sideband}
\end{figure}

\begin{figure}[p]
  \centering
    \includegraphics[width=4in]{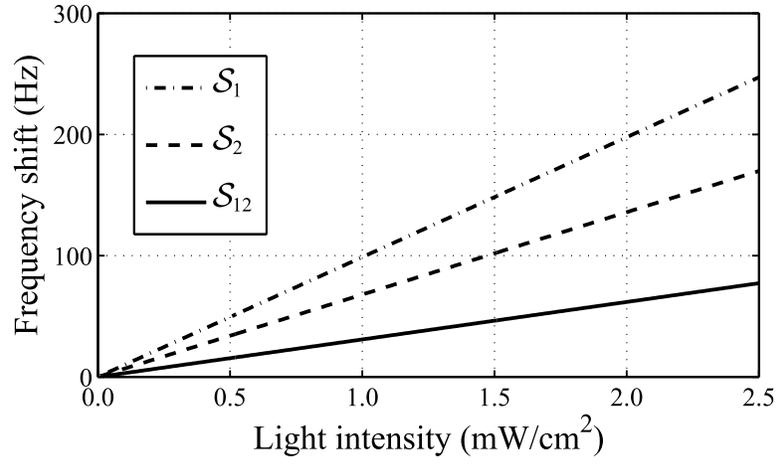}
  \caption{Light shift as a function of light intensity.}
  \label{fig:light shift_CW}
\end{figure}

\end{document}